\documentclass[usenatbib, usegraphicx]{mn2e}

\title[]{Discovery of an overdensity of faint red galaxies in the vicinity of  the $\mathbf{z=1.786}$ radio galaxy 3C~294\thanks{Based on observations made with ESO Telescopes at the Paranal Observatory under programme ID 68.A--0560}}

\author[S. Toft et al.]
       {S. Toft,$^1$\thanks{E-mail: toft@astro.ku.dk},
	 K. Pedersen,$^1$, H. Ebeling$^2$ and J. Hjorth$^1$ \\
 $^1$Astronomical Observatory, University of Copenhagen, Juliane Maries Vej 30, 
DK-2100 Copenhagen \O, Denmark \\ $^2$ Institute for Astronomy, University of Hawaii, 2680 Woodlawn Drive, Honolulu, HI96822, USA }

\pagerange{\pageref{firstpage}--\pageref{lastpage}}
\pubyear{2003}

\newcommand{\beas}{\begin{eqnarray*}}
\newcommand{\eeas}{\end{eqnarray*}}
\newcommand{\mnras}{MNRAS}
\newcommand{\apj}{ApJ}

\newcommand{\apjl}{ApJL}

\newcommand{\aap}{A\&A}
\newcommand{\aj}{AJ}

\usepackage{natbib}
\bibdata{3c294letter}

\begin{document}

\maketitle
\label{firstpage}

\begin{abstract}
We report the discovery of an overdensity of faint red galaxies in the vicinity of the $z=1.786$ radio galaxy 3C~294. The overdensity, discovered in a 84 min $Ks$-band ISAAC/VLT image is significant at the $2.4\sigma$ level (compared to the local field density), and overlaps with the extended X-ray emission around 3C~294 detected with the Chandra X-ray Observatory.
The near-infrared colours of the galaxies making up the overdensity show a large scatter and the galaxies do not follow a red sequence in the colour magnitude diagram. If the galaxies are in a cluster at $z=1.786$ they must be dominated by young stellar populations with different star-formation histories.

\end{abstract}


\section{introduction}
The most powerful low-redshift radio galaxies are giant elliptical galaxies and are often found to be dominant central galaxies in rich clusters.
If this is also the case at high redshift, luminous radio galaxies may be useful tracers of distant clusters.
\cite{fabian01} reported the detection of diffuse X-ray emission with the characteristics of an 
intracluster medium, surrounding the $z=1.786$ radio galaxy 3C~294, suggesting the presence 
of a cluster of galaxies around it. However, the 200~ksec Chandra follow-up
study (\cite{fabian03}) indicates that the diffuse emission is most likely non-thermal, but confined by an intracluster medium.
3C~294 is a  radio source associated with an emission-line galaxy, with a double-hotspot 
radio morphology suggestive of precessing jets originating from the weak flat-spectrum core 
\citep{mccarthy90}.
The galaxy is embedded in a elongated Ly~{$\alpha$} nebula roughly aligned with the radio 
structure \citep{mccarthy90}.

A bright ($V=11\fm5$) star $11\arcsec$ west of 3C~294 has hampered deep optical observations of 
the surrounding field, but has allowed the use of adaptive optics for deep, high-resolution imaging of the radio galaxy at near-infrared (NIR) wavelengths \citep{stockton1999,quirrenbach2001}. 
These observations  resolve the radio galaxy 
in several distinct subclumps in the process of merging, supporting the idea that 3C~294 is located 
in a dense environment.   
In this Letter we present NIR imaging of the field around 3C~294 revealing intriguing evidence for an overdensity of faint red galaxies in the vicinity of 3C~294. 
Sec.~\ref{data} describes the observations, the data reduction and analysis techniques. In Sec.~\ref{results} we present 
results on the spatial and colour distribution of galaxies in the field, 
and in Sec.~\ref{discussion} we discuss the results.  

Throughout this Letter we assume a flat $\Omega_m=0.3$,
$\Omega_{\Lambda}=0.7$, $h=0.65$ cosmology. At a redshift of 1.786, one arcsec corresponds to
an angular distance of 9.1~kpc.

\section{Data}
\label{data}
We have obtained NIR observations at the field around 3C~294 with the ISAAC 
instrument on the VLT UT1 telescope. The observations were carried out in service mode on April 1, 2002 and consist of 84 min in the $Ks$ band and 40 min in each of the $H$ 
and $Js$ bands. The exposure time at the individual 
dither positions was kept short (10 sec in all bands) to reduce the effects of the bright star $11\arcsec$ 
west of 3C~294 on the surrounding field. 

The data were reduced using a combination of the {\emph{eclipse}} software package \citep{eclipse} 
and the IRAF ``Experimental Deep Infrared Mosaicing
Software'' {\emph{xdimsum}}. {\emph{Eclipse}} was used to remove effects of electrical ghosts
from science and calibration frames and to construct flat fields and bad-pixel maps from a 
series of twilight sky flats. 
Sky subtraction and combination of the individual science frames was carried out with 
{\emph{xdimsum}}.
The $H$ and $Js$ band images were geometrically transformed to the $Ks$ band image 
(using the IRAF tasks {\emph{geoform} and {\emph{geotran}}).
The observations were performed in photometric weather conditions. The seeing in the combined 
frames is approximately $1\farcs1$ in all wavebands.

For the photometric calibration we used zero-points derived for that night by the ESO staff (corrected for atmospheric extinction). These are accurate to about $0\fm06$ and can be found on the ESO webpage\footnote{http://www.eso.org/observing/dfo/quality/ISAAC/img/trend/}.

Object detection was performed in the $Ks$ band, using the {\emph{SExtractor}}
software \citep{bertin} with default parameter settings.
We searched for the detected objects in the remaining frames using the
{\emph{SExtractor} ASSOC option.
The SExtractor {\emph{class\_star}} parameter was used to exclude obvious stars (objects with well-defined FWHM, and  {\emph{class\_star}} $>$ 0.5) from the sample.
For deriving the total $Ks$ band magnitudes of the objects we used  the
{\emph{SExtractor} parameter
{\emph{mag\_best}} which has been shown to be accurate to about
$0\fm05$ \citep{smail2001}. Colours were derived in fixed apertures
of 10 pixels which corresponds to approximately 13 kpc at $z=1.786$ with the ISAAC pixel scale of $0\farcs147$ pix$^{-1}$.

\begin{figure}
\setcounter{figure}{0}
\resizebox{\hsize}{!}{{\includegraphics{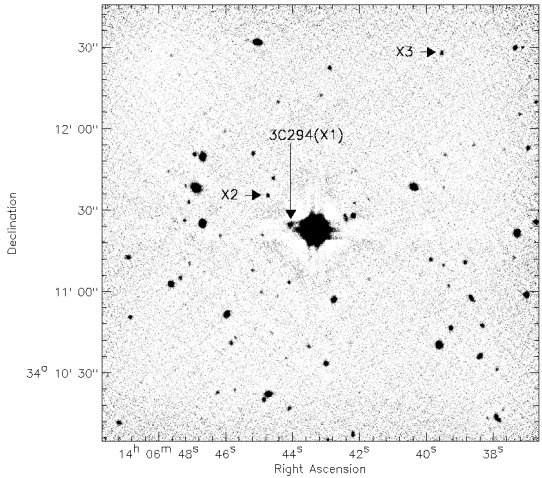}}}
\caption{ISAAC/VLT $Ks$-band image of the field around 3C~294. 
The $Ks$ band counterparts of the three brightest X-ray 
point sources (see Fig.~\ref{pointsources}) are indicated by arrows.}
\label{kfullfield}
\end{figure}

\begin{figure*}
\setcounter{figure}{1}
\resizebox{\hsize}{!}{{\includegraphics{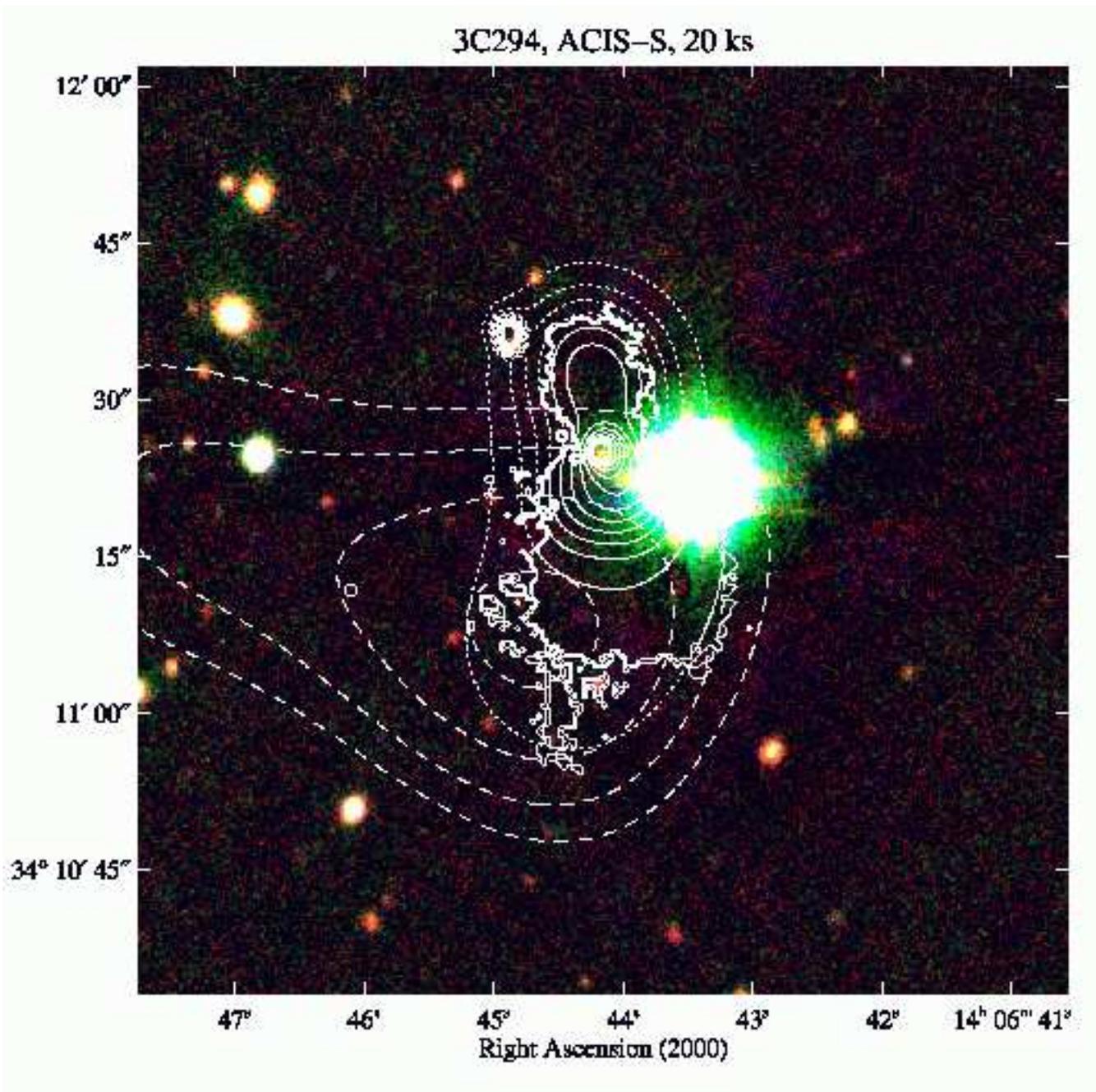}}}
\caption{Composite $Ks$, $H$, $Js$ band image of the central 2$\arcmin \times 2\arcmin$ field around 3C~294 with full exposure time. 
The dashed lines show contours of the adaptively smoothed density distribution of faint ($Ks > Ks_{(\rm{3C~294})}$) galaxies. Contours start at 2.5 times the local background density and increase by 5\% between adjacent contours.   
The dotted and solid lines show isointensity contours of the adaptively
 smoothed X-ray emission as observed with ACIS-S. Contour levels start 20\%
 above the global background value and increase by 20\% between adjacent
 contours. Solid lines within the ragged, bold contour delineate features
 found to be more than $3\sigma$ significant compared to the local background;
 X-ray contours outside this region correspond to emission of lower
 significance and are plotted as dotted lines.
 }
\label{kcolzoom}
\end{figure*}

A 20~ksec observation with the ACIS-S detector aboard the Chandra X-ray Observatory was obtained on October 29, 2000 \citep{fabian01} with 3C~294 placed near the nominal aim point on the back-illuminated S3 CCD. Our analysis uses the standard data products from the public Chandra archive. 

We aligned the NIR and CXO images by tying them separately to the Second STScI Digitized Sky Survey  image of the field (DSS-II, e.g. \cite{mclean2000}), in order to avoid having to assume that the X-ray and optical emission from 3c294 originate in the same spot.

\section{Results}
\label{results}
In Fig.~\ref{kfullfield} we show the $Ks$ band image of the field around 3C~294. The short exposure time and careful background subtraction confines the extent of the bright star to a circular area of radius $\sim 7\arcsec$ and a  surrounding circular area (extending to radius $\sim 14\arcsec$) where the background is slightly oversubtracted.
The derived  photometric errors of objects in this area are not significantly larger than those derived for objects further away from the star. 
The derived $Ks$-band magnitude of 3C~294 (see Table~\ref{table}) is consistent with the magnitude $K=18.0\pm0.3$ found by \cite{mccarthy90} in a 3{$\arcsec$} aperture.

\subsection{X-ray emission}
Both pointlike and extended diffuse X-ray emission are detected 
in the direction of 3C~294.
We show the adaptively smoothed X-ray emission as detected in this
ACIS-S observation in  Fig.~\ref{kcolzoom}. Using the {\sc asmooth} algorithm of \cite{ebeling2003} we adjust the scale of the
Gaussian smoothing kernel such that all features apparent in the
smoothed X-ray image are at least $3\sigma$ significant. This
significance criterion is met within the ragged, bold contours
shown in  Fig.~\ref{kcolzoom}. As noted already by Fabian and coworkers the
diffuse emission is very compact (about 100 kpc in radius at the
redshift of 3C~294) and hour-glass shaped.

\subsection{Distribution of faint galaxies}
In addition to a few dozen bright foreground galaxies, a substantial number of faint red galaxies were detected in the immediate vicinity of the radio galaxy. The significance of this overdensity was quantified from a smoothed galaxy density map of the central $2\arcmin \times 2\arcmin$ part of the field. Due to the dithering pattern, only the central $2\arcmin \times 2\arcmin$ had the full exposure time.
An image was constructed where the pixel values at the centroid of faint ($Ks > Ks_{(3C~294)}$) galaxies were set equal to 1 and the rest of pixels in the image were set to zero. This image was then adaptively smoothed, using the  {\sc asmooth} algorithm of \cite{ebeling2003}, aiming at $3\sigma$ significance.  
In Fig.~\ref{kcolzoom} we superpose the contours of the resulting smoothed galaxy density distribution and the contours of the Chandra observations of the diffuse X-ray emission on a composite a  $Js$, $H$, $Ks$ image of the central $2\arcmin \times 2\arcmin$ of the field.
The galaxy density contours start at 2.5 times the local background density and increase by 5\% between the adjacent contours. The local background density was estimated as the mean density of galaxies in the field (excluding the part of the image containing the overdensity).   From Fig.~\ref{kcolzoom} there is evidence for an overdensity of faint galaxies in the vicinity (within 150 kpc) of 3C 294. The peak significance reached at the center of the contours is 2.4~$\sigma$. The contours of the galaxy overdensity overlap with the contours of the extended X-ray emission but are more extended.
It should be noted that a substantial number of faint galaxies that could increase the significance of the overdensity are likely to be hidden behind the star and in the surrounding area with poor sky subtraction.

\subsection{Colours of the faint galaxies}
In Fig.~\ref{JK} we plot $Js-Ks$ vs $Ks$ for faint ($Ks > Ks_{(\rm{3C~294})}$) galaxies in the vicinity of 3C~294. Also indicated is the predicted  colour magnitude relation at the redshift of 3C~294 for populations of passively evolving elliptical galaxies formed at redshifts $z_f=2$, 3 and 5 \citep{kodama97}.
No distinct ``red sequence'' \citep{SED98} is  seen in the diagram. The red sequence of cluster galaxies is predicted to increase its scatter and eventually fall apart as the redshift approaches the formation redshift of the stars in the galaxies.
If the majority of the galaxies that make up the overdensity are at the redshift of the radio galaxy, the absence of a red sequence may be due to differences in the dust content, the morphology and/or the age of the stellar populations of the  galaxies.
Alternatively, the galaxies that make up the overdensity could be a chance alignment of galaxies at different redshifts.  

3C~294 has been observed with the WFPC2 camera (aboard the Hubble Space Telescope) through the F702W filter in two epochs. These data are however too shallow  to detect the galaxies that make up the overdensity (with a total exposure time of 19 min, the radio galaxy is only marginally detected).

\begin{figure}
\setcounter{figure}{2}
\resizebox{\hsize}{!}{{\includegraphics{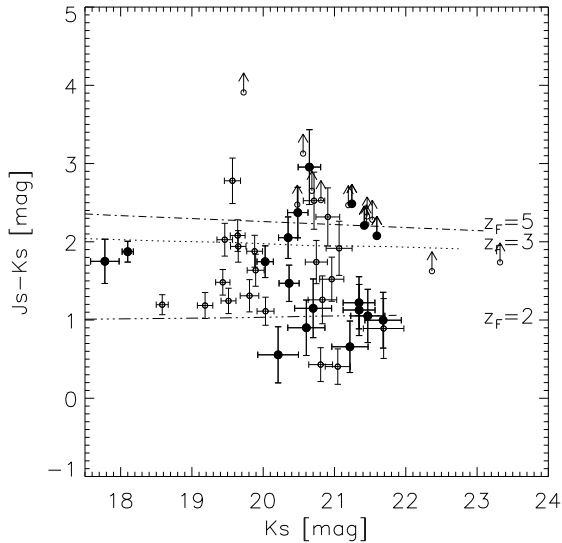}}}
\caption{$Js-Ks$ vs.\ $Ks$ for faint ($Ks > Ks_{(\rm{3C~294})}$) galaxies in the vicinity of 3C~294.  Large filled symbols represent galaxies within 25{$\arcsec$} of the overdensity peak. Small open symbols represent galaxies  within 50{$\arcsec$} of the overdensity peak. Galaxies detected only in $Ks$ are assigned $5\sigma$ upper limit $Js$ band magnitudes. 
Error bars represent statistical errors from {\emph{SExtractor}} and photometric zero-point errors added in quadrature. 
The three lines are  predicted  colour-magnitude relations at the redshift of 3C~294 ($z=1.786$) of a population of passively evolving elliptical galaxies formed at $z_f=2$, 3 and 5 \citep{kodama97}   }
\label{JK}
\end{figure}

\subsection{X-ray point sources}
Consistent with expectations from deep Chandra observations \citep{rosati2002} of about three random X-ray point 
sources within the field covered by the $Ks$-band image, we detect 3C~294 and two other X-ray point sources in the Chandra 0.5--5~keV image.
Properties of these point sources are given in  Table~\ref{table}.
They all have obvious, extended NIR counterparts (see Fig.~\ref{pointsources})
with $Js-Ks$ colours compatible with colours of the faint galaxies
around 3C~294. 
If the sources are situated at the redshift
of 3C~294, their X-ray luminosities are typical of active galaxies. 
The derived hardness ratios of the point sources suggest that X2 is an AGN type II, and that X3 is an AGN type I (Fig.~3 in \cite{rosati2002}). 
In a Chandra study of the nearby galaxy cluster Abell 2104,
\cite{martini2002} detect two active galaxies with $L_X>10^{43}$~erg~s$^{-1}$
(the detection limit of the 3C~294 Chandra image). 
The properties of the X-ray point sources in the field of 3C~294 
are thus consistent with the expectation for a cluster at $z=1.786$.

\begin{table}
\caption{X-ray luminosities ($L_X$), Hardness Ratios (HR), NIR magnitudes and NIR positions of the NIR 
counterparts of the X-ray point sources in the field of 3C~294. $L_X$ is
given in the $z=1.786$ restframe 0.5--10~keV band in units of $10^{44}$~erg~s$^{-1}$, corrected for Galactic absorption (and for 3C~294 also for intrinsic absorption) assuming a power law spectrum with a photon index of 1.5. The hardness ratio is defined as HR=(H$-$S)/(H+S), where S is the net number of counts in the 0.5--2~keV band and H is the net number of counts in 2--8~keV band. The error on the derived hardness ratios is 0.2.   
}
\label{table}

\begin{tabular}{|lrrr|}\hline
      & 3C~294 (X1)   & X2		& X3			\\ \hline
$L_X$ & 6.9                & 1.8               & 0.8\\
HR    & 0.6	      &  $-0.1$	&	$-0.6$ \\ 	
$Ks$  & $17.78\pm0.07$ &$19.64\pm0.10$ & $19.33\pm0.09$ \\
$H-Ks$&  $0.86\pm0.28$ & $0.65\pm0.15$ & $0.96\pm0.15$ \\
$Js-Ks$& $1.75\pm0.31$ & $2.08\pm0.20$ & $2.26\pm0.20$ \\ 
RA(J2000.0)    &	14:06:44.0 & 14:06:44.7 & 14:06:39.5 \\ 
Dec(J2000.0)   &+34:11:25.4 &+34:11:36.4 & +34:12:29.6  \\ \hline
\end{tabular}
{\small{}}
\end{table}

\begin{figure}
\setcounter{figure}{3}
\resizebox{\hsize}{!}{{\fbox{\includegraphics{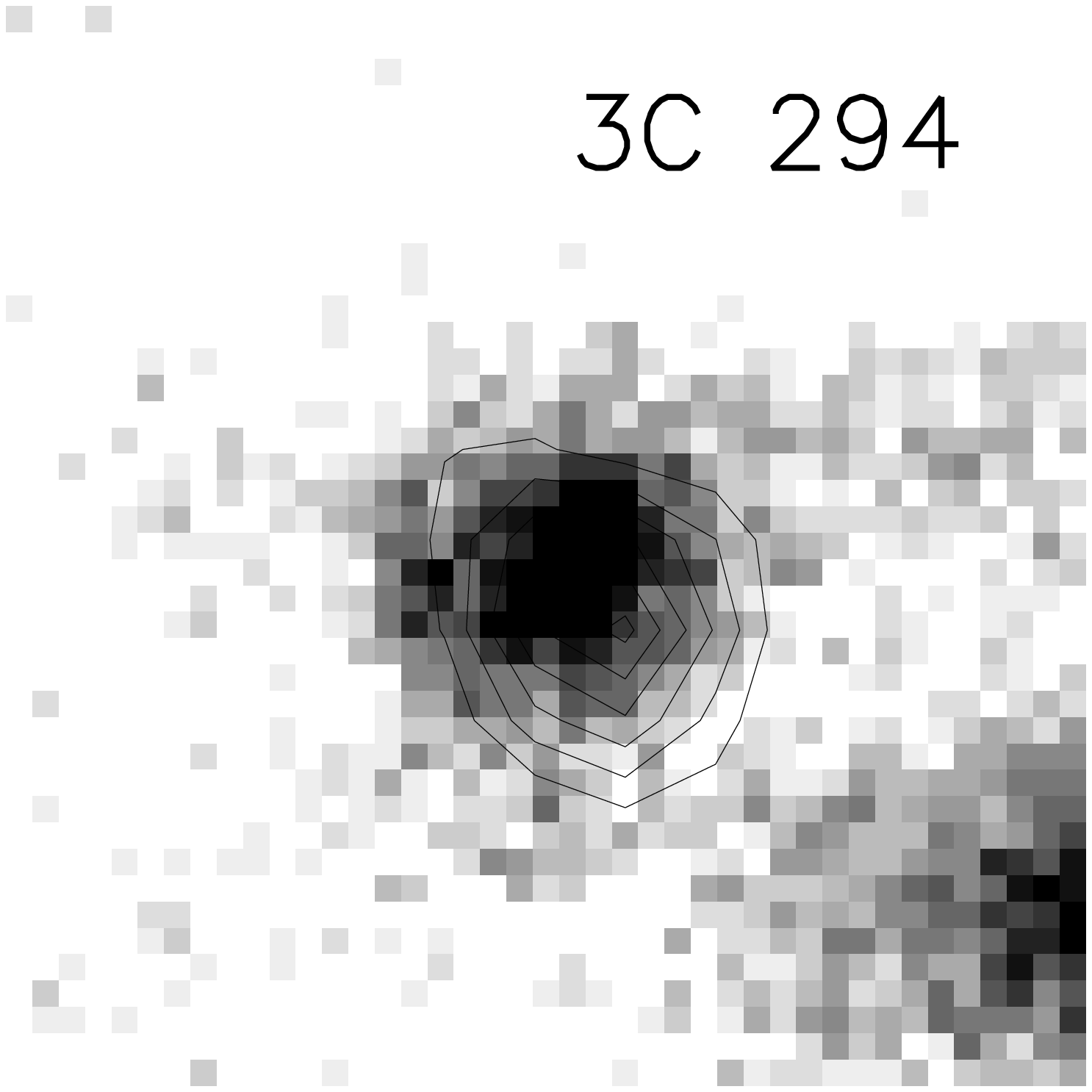}},\fbox{\includegraphics{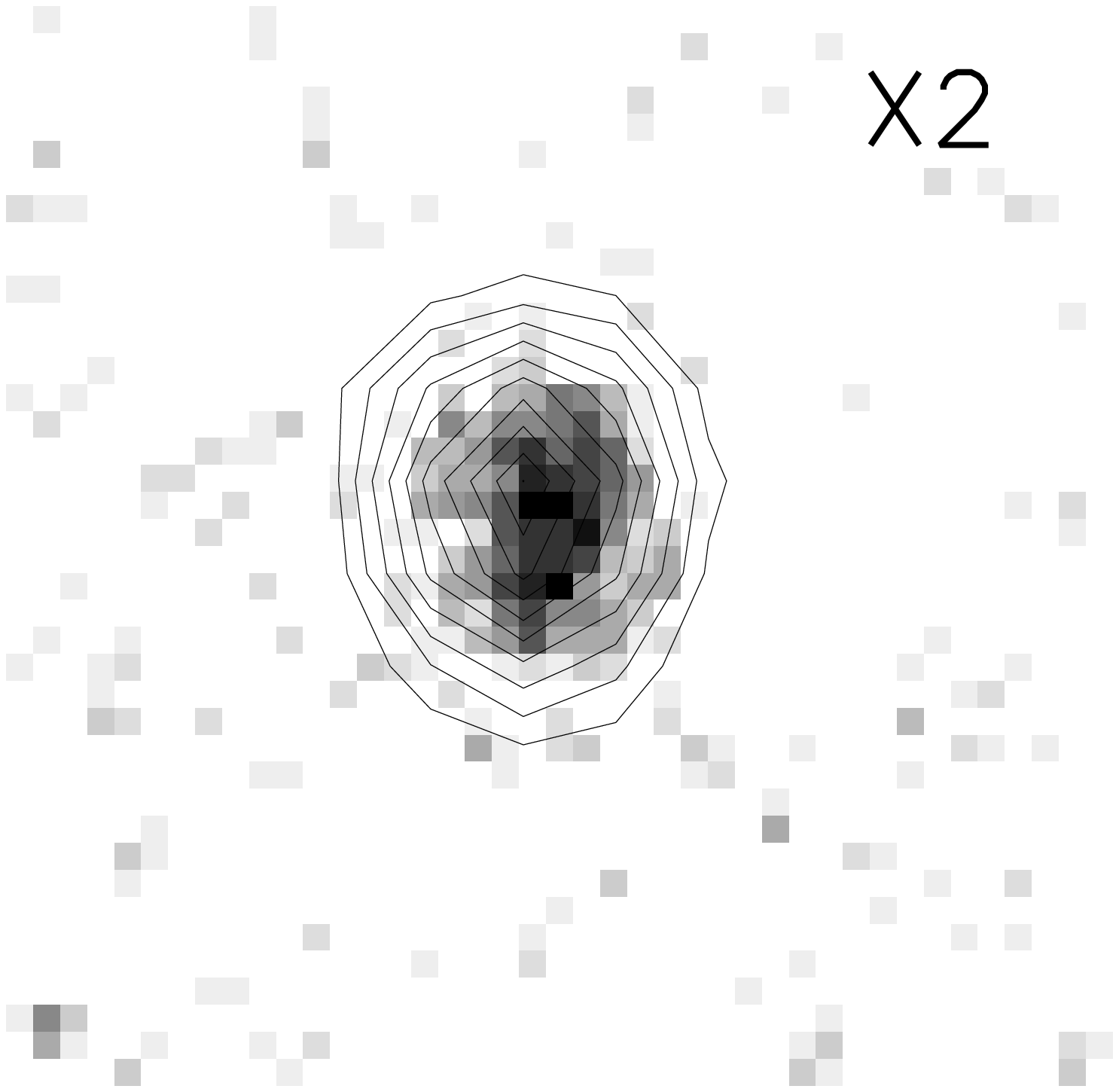}},\fbox{\includegraphics{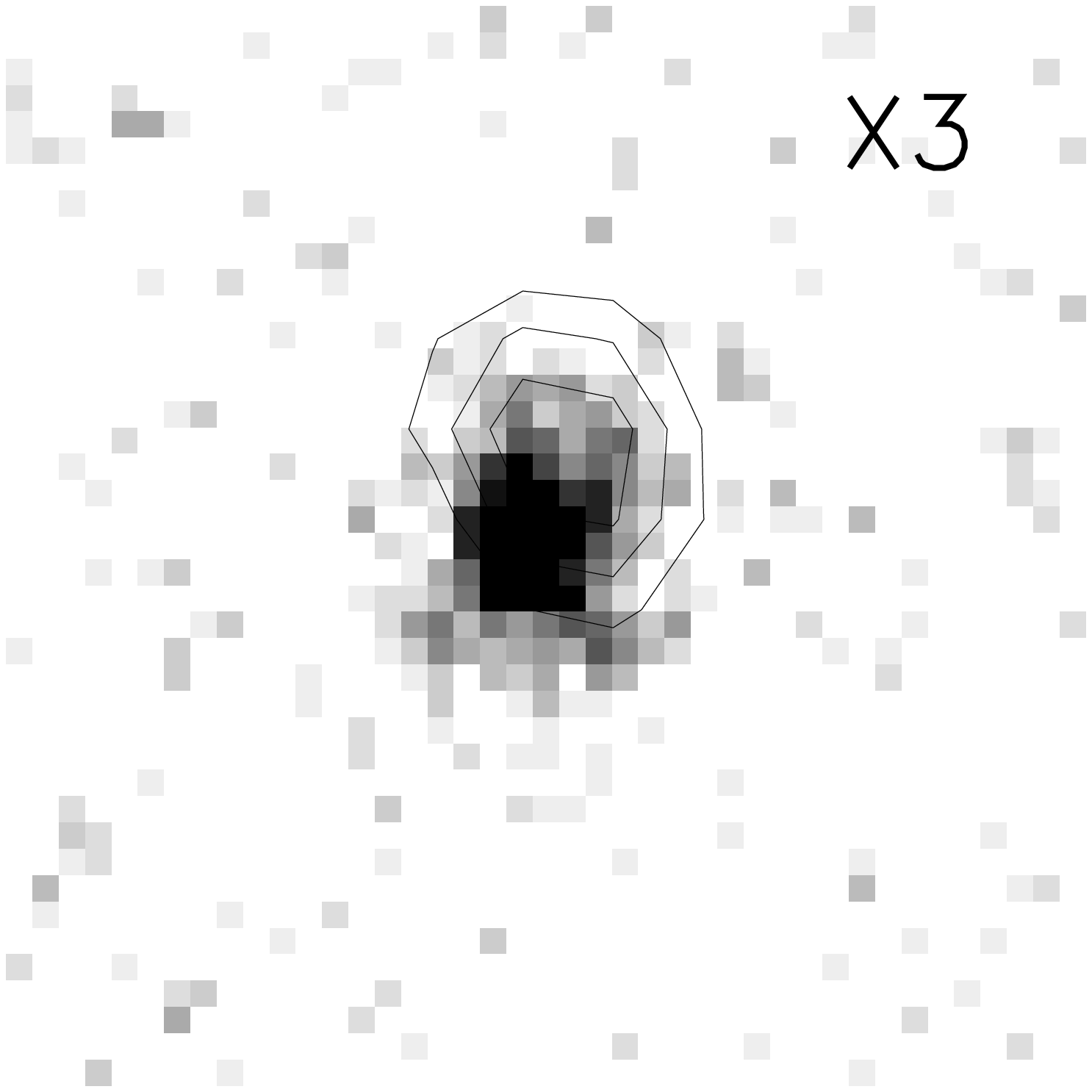}}}}
\caption{$54 \times 54$ kpc$^2$ ``postage stamps'' of the $Ks$ band identifications of the three X-ray point sources in the field (represented by contours)  }
\label{pointsources}
\end{figure}

\section{Discussion}
\label{discussion}

We have presented evidence for an overdensity of faint red galaxies in the vicinity of the massive $z=1.786$ radio galaxy 3C~294, which overlaps with the extended soft X-ray emission detected with Chandra.
We note, however, that the misalignment of the X-ray and galaxy density
 centroids as well as their very different degrees of angular extent and
 concentration makes it unlikely that the X-ray emission is due to a relaxed intracluster medium.  

The galaxy overdensity detection and the X-ray emission are, as independent pieces of evidence for a $z=1.786$ cluster in the field, only marginally significant. 
However, considering that they overlap with each other and with the position of a massive radio galaxy (believed to trace high density environments) makes the case for a cluster of galaxies around 3C~294 appealing, although it has to be confirmed by additional deeper, more detailed studies.

At this redshift, clusters are expected to be in a transition phase between that of the $z\ga2$ proto clusters picked out by Ly$\alpha$ imaging techniques \citep{pentericci2000} and the massive $z\la1.3$ clusters selected by optical or X-ray techniques \citep{SED98,rosati99,ebeling2000,ebeling2001}, and therefore very valuable both for galaxy evolution studies, and structure evolution studies in general.

Studies of the galaxy population, even in the highest redshift clusters known to date, have not been able to distinguish between the hierarchical formation scenario which is supported by an  increasing fraction of blue and  merging galaxies with redshift \citep{vandokkum00,vandokkum01} and the monolithic elliptical galaxy formation scenario \citep{eggen} which is supported by the presence of the well defined ``red sequence'' of elliptical galaxies in most known clusters \citep{gladders98}. The presence of the red sequence favors  monolithic collapse, but does not exclude the possibility that cluster ellipticals formed from mergers of smaller galaxies as long as the bulk of star formation took place at much larger redshifts and there was little star formation in the subsequent merging process.

If the galaxies making up the overdensity presented here are confirmed to be in a cluster at $z=1.786$ it will be the first example of a cluster where the characteristic red sequence has not yet formed and the colours of the galaxies are dominated by young stellar populations with different star formation histories. Such a system is potentially very powerful for constraining galaxy formation and evolution models.

\label{lastpage}

\section{Acknowledgments}
We thank T. Kodama for providing us with his elliptical galaxy evolution models, N. Drory and G. Feulner for letting us use their scripts for distortion corrections. ST acknowledges support from the Danish Ground-Based Astronomical Instrument Centre (IJAF). This work was supported by the Danish Natural Science Research Council (SNF).

\bibliography{3c294letter}

\end{document}